\documentclass[cits]{PoS}

\def\rhodrat{\rho_\mathrm{d}/\rho_\mathrm{h}}

\def\GeV{GeV$/$c$^2$}

\title{Signatures of the Milky Way's Dark Disk in Current and Future Experiments}

\ShortTitle{Signatures of the Milky Way's Dark Disk in Current and Future Experiments}

\author{\speaker{T. Bruch}\thanks{Physics Institute} , J. Read\thanks{Institute for Theoretical Physics} , L. Baudis$^\dagger$ and G. Lake$^\ddagger$\\
        University of Zurich, Winterthurerstrasse 190 8047 Zurich, CH\\
        E-mail: \email{[tbruch justin lbaudis lake]@physik.uzh.ch}}

\abstract{In hierarchical structure formation models of disk galaxies, a dark matter disk forms as massive satellites are preferentially dragged into the disk-plane where they dissolve. Here, we quantify the importance of this dark disk for direct and indirect dark matter detection. 

The low velocity of the dark disk with respect to the Earth enhances detection rates in direct detection experiments at low recoil energy. For WIMP masses $M_{WIMP}\gtrsim50$\,\GeV, the detection rate increases by up to a factor of 3 in the $5 - 20$\,keV recoil energy range. Comparing this with rates at higher energy is sensitive to $M_{WIMP}$, providing stronger mass constraints particularly for $M_{WIMP}\gtrsim100$\,\GeV. The annual modulation signal is significantly boosted by the dark disk and the modulation phase is shifted by $\sim3$ weeks relative to the dark halo. The variation of the observed phase with recoil energy determines $M_{WIMP}$, once the dark disk properties are fixed by future astronomical surveys. 

The low velocity of the particles in the dark disk with respect to the solar system significantly enhances the capture rate of WIMPs in the Sun, leading to an increased flux of neutrinos from the Sun which could be detected in current and future neutrino telescopes. The dark disk contribution to the muon flux from neutrino back conversion at the Earth is increased by a factor of $\sim5$ compared to the SHM, for $\rhodrat=0.5$.  }

\FullConference{Identification of dark matter 2008\\
		 August 18-22, 2008\\
		 Stockholm, Sweden}

\begin{document}

\section{Introduction}
 Recent  $\Lambda$CDM simulations of galaxy formation including the effect of the baryons demonstrated that massive satellites are preferentially dragged into the baryonic disk plane by dynamical friction where they dissolve, leaving a thick dark matter disk \cite{Read08}. Here we study the effects of the dark disk on current and future direct and indirect detection experiments. We consider a dark disk with density ratios to the Standard Halo Model (SHM) density ($\rho_h = 0.3$\,GeV/cm$^3$) of $\rhodrat=[0.5,1,2]$. We assume that the dark disk kinematics match the Milky Way's stellar thick disk. At the solar neighborhood, this gives a rotation lag $v_{lag}$ of $40-50$\,km/s with respect to the local circular velocity, and dispersions of $(\sigma_R,\sigma_\phi,\sigma_z) = (63,39,39)$\,km/s \cite{Read08}. Since the dispersions are nearly isotropic and somewhat uncertain, we model the velocity distribution of particles in the dark disk and the SHM with a simple 1D Maxwellian:
\begin{equation}
f(\textbf{v},t) \propto \exp\left(\frac{-(\textbf{v}+\textbf{v}_\oplus(t))^2}{2\sigma^2}\right)
\end{equation}
\noindent
where $\textbf{v}$ is the laboratory velocity of the dark matter particle and the instantaneous streaming velocity
$\textbf{v}_\oplus=\textbf{v}_\mathrm{circ}+\textbf{v}_\odot+\textbf{v}_\mathrm{orb}(t)$. This
streaming velocity is the sum of local circular velocity $\textbf{v}_\mathrm{circ} =(0, 220, 0)$\, km/s , the peculiar motion of the Sun $\textbf{v}_\odot = (10.0, 5.25, 7.17)$\, km/s  \cite{DehnenBinney} with respect to $\textbf{v}_\mathrm{circ}$
and the orbital velocity of the Earth around the Sun $\textbf{v}_\mathrm{orb}(t)$. In the SHM, the halo has no rotation and the dispersion $\sigma=|\textbf{v}_\mathrm{circ}|/\sqrt{2}$. For the dark disk, the velocity lag $\textbf{v}_{lag} =(0, 50, 0)$\, km/s replaces $\textbf{v}_\mathrm{circ}$ and a dispersion of $50$\, km/s is adopted.

\begin{figure*}
\begin{center}
\includegraphics[width=0.48\textwidth]{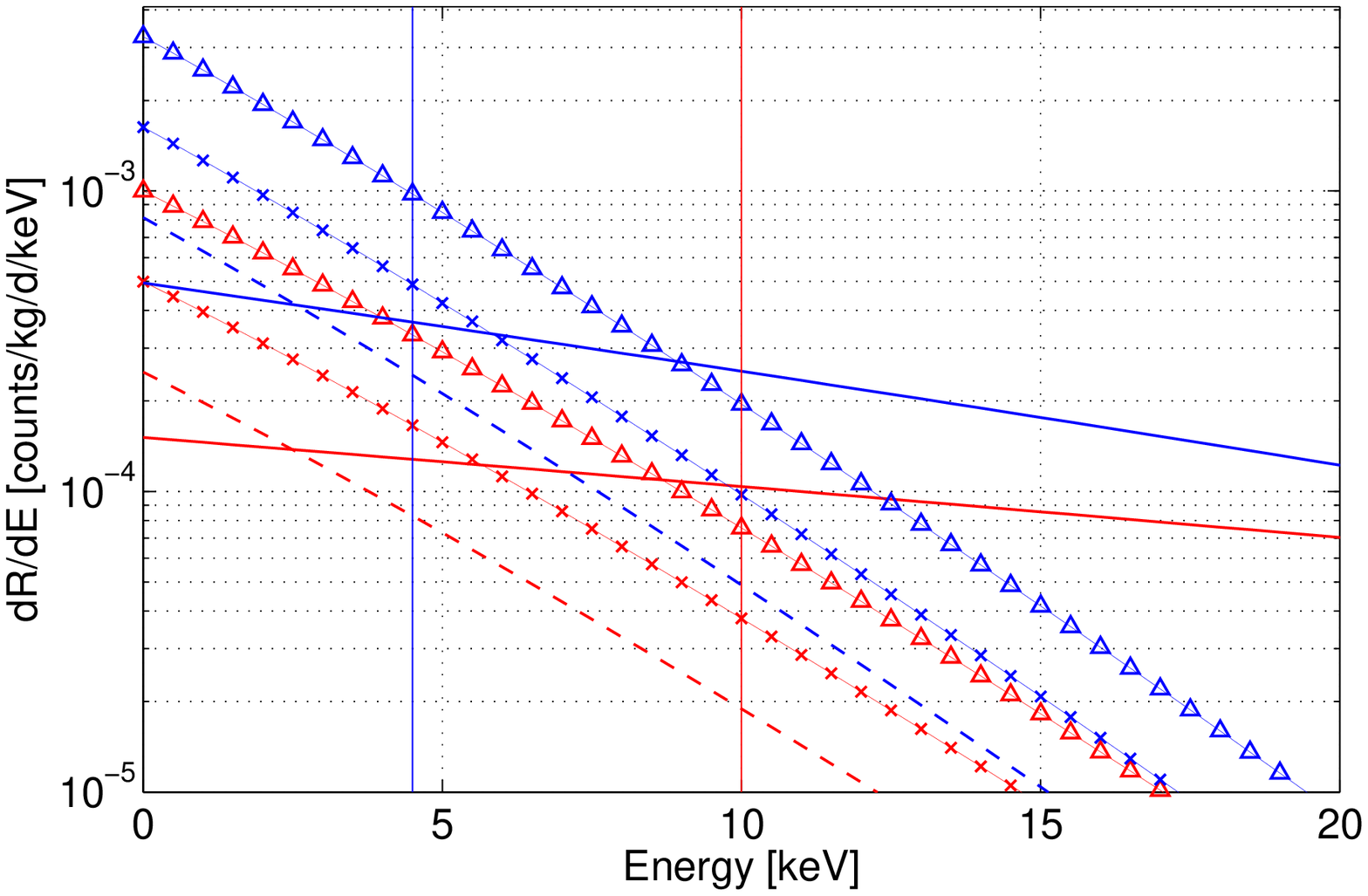}
\includegraphics[width=0.48\textwidth]{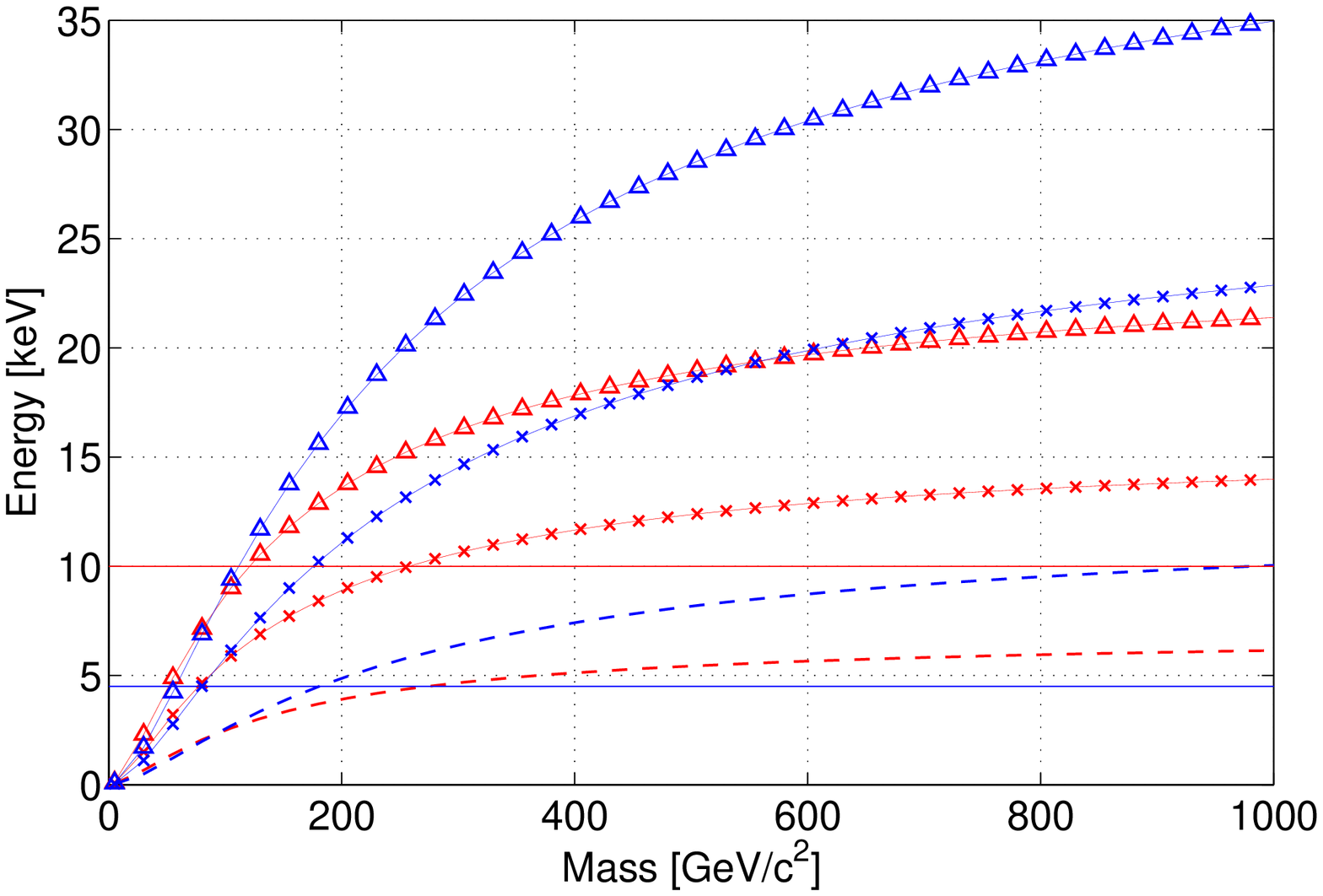}

\caption{Left panel: Differential recoil rates for Ge (red) and Xe (blue) targets, for WIMPs with $M_{WIMP}=100$\, GeV and a WIMP nucleon cross section of 10$^{-8}$ pb in the SHM (solid line) and the dark disk. Right panel: The recoil energy below which the signal is dominated by the dark disk (compared to the SHM) as a function of $M_{WIMP}$ for  Ge (red) and Xe (blue) targets. Three different values of $\rhodrat$ (0.5 dashed, 1 {\large{$\times$}} and 2 $\bigtriangleup$) are shown. Vertical / horizontal lines mark current experiment thresholds: XENON10 \cite{Xenon} (blue) using a Xe and CDMS-II \cite{CDMS} (red) using a Ge target.}
\end{center}
\end{figure*}

\section{Direct Detection Experiments}
 Direct detection experiments measure nuclear recoils above threshold to detect WIMPs via their elastic scattering on target media; here we consider Ge and Xe. We consider only the spin-independent scalar WIMP-nucleus coupling since it dominates the interaction for target media with nucleon number A $\gtrsim$ 30. The expected recoil rate per unit time, unit mass, unit nuclear recoil energy and unit time is given by:
\begin{equation}
\frac{dR}{dE} = \frac{\rho \sigma_\mathrm{wn} |F(E)|^2}{2 M_{WIMP} \mu^2}  \int_{v>\sqrt{m E/ 2 \mu^2}}^{v_{max}} \frac{f(\textbf{v},t)}{v} d^3v
\end{equation}
\noindent
where $\rho$ is the local dark matter density, $\sigma_\mathrm{wn}$ is the WIMP-nucleus scattering cross section, $F(E)$ is the nuclear form factor,  
$M_{WIMP}$ and m are the masses of the dark matter particle and of the target nucleus, respectively, 
$\mu$ is the reduced mass of the WIMP-nucleus system, $v=|\textbf{v}|$ and $v_{max}$ is the maximal velocity in the earth frame for particles moving at the galactic escape velocity $v_{esc}=544$\,km/s \cite{Rave07}.

 Since the lower relative velocity of the dark disk significantly increases the differential recoil rate at low energies compared to the SHM rate (Fig.1, left panel), the detection of the dark disk depends crucially on the detector's low energy threshold. The energy below which the dark disk contribution dominates the differential rate as a function of $M_{WIMP}$ is shown in Fig.1 (right panel) for three values of $\rhodrat$. The total rate in a detector is the sum of the dark disk and the SHM contribution. For $M_{WIMP}$ $\gtrsim$ 50\,\GeV, the dark disk contribution lies above current detector thresholds. The details of the summed differential rate with energy betrays both the contribution of the dark disk relative to the SHM and $M_{WIMP}$. This introduces a mass dependent characteristic shape of the differential rate that will improve the constraints on $M_{WIMP}$ upon detection \cite{darkdisk}.

\begin{figure*}
\begin{center}
 \includegraphics[width=0.32\textwidth,height=39mm]{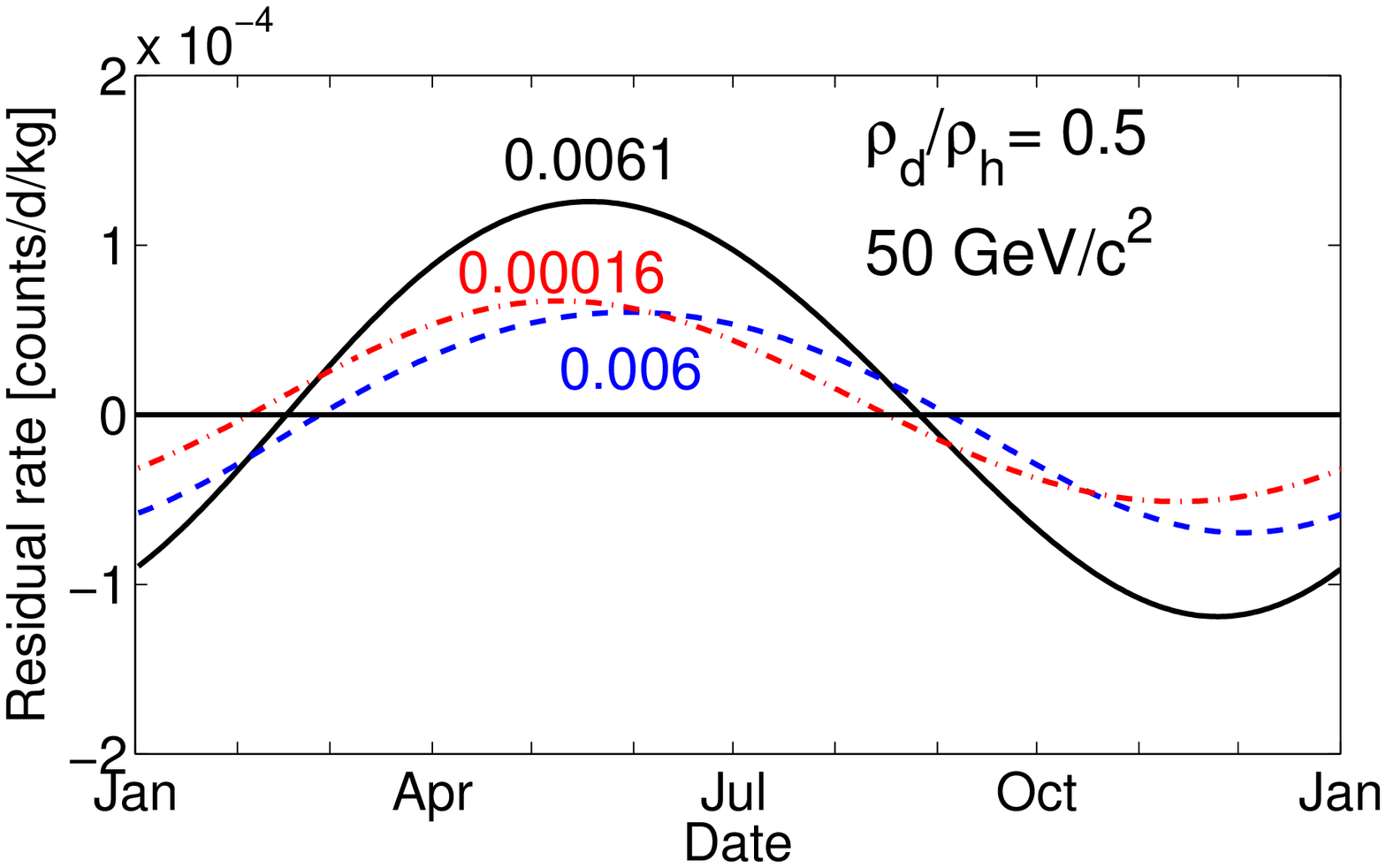} 
 \includegraphics[width=0.32\textwidth,height=39mm]{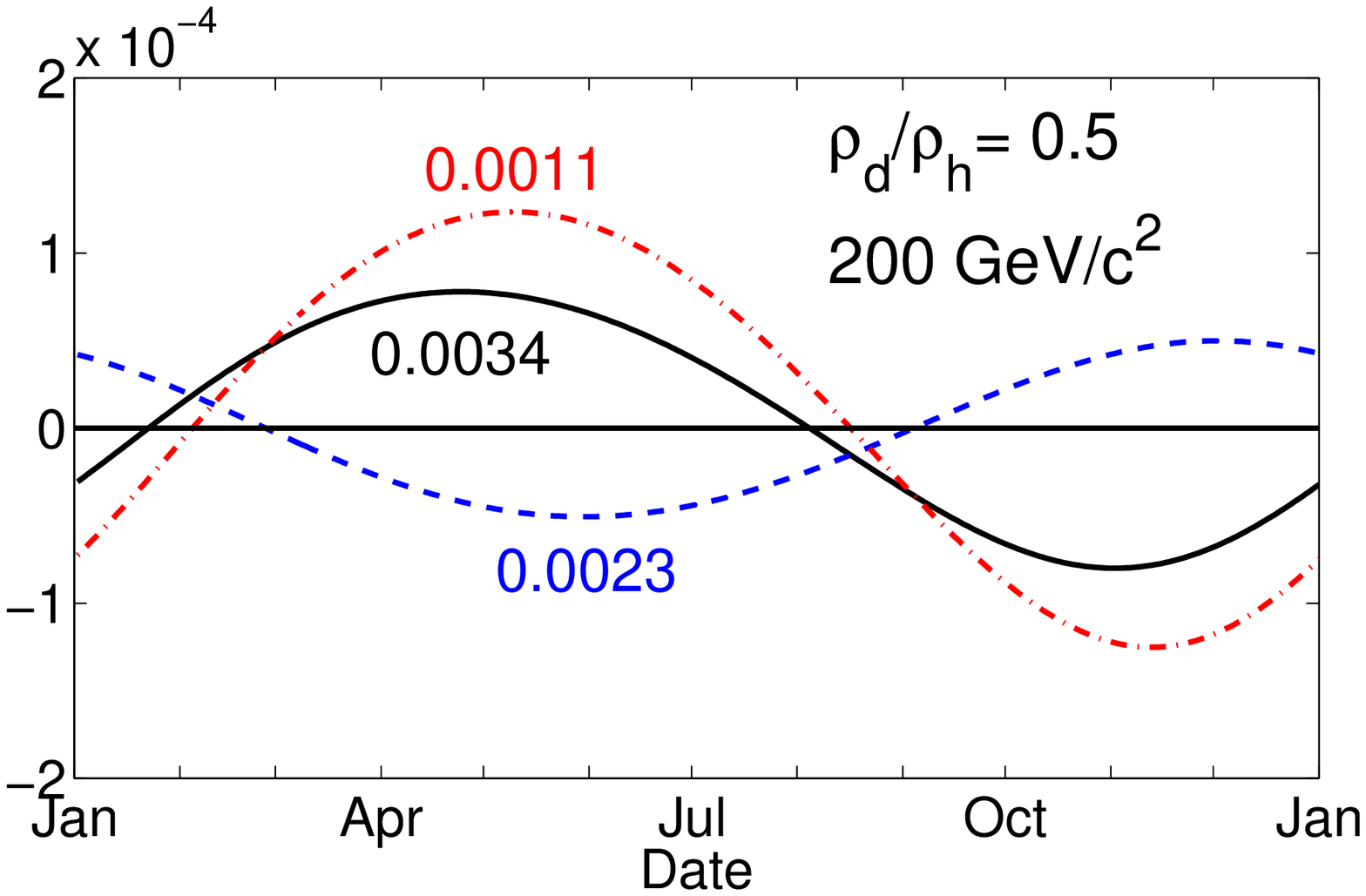} 
 \includegraphics[width=0.32\textwidth,height=39mm]{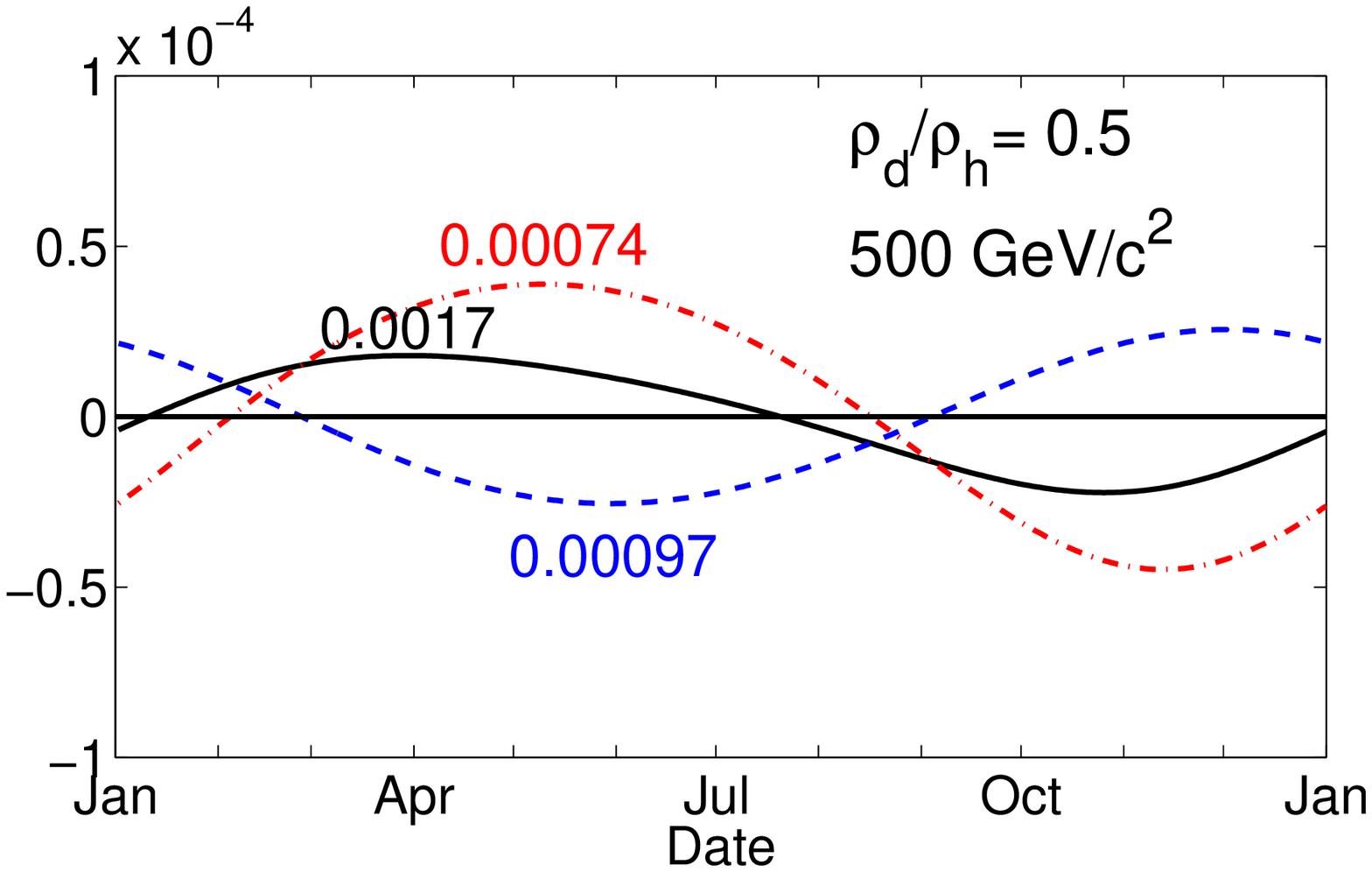} \\
\end{center}
\caption{The annual modulation shown as the residual counting rate versus date for the XENON10 experiment (4.5 to 27 keV). The residuals are calculated with respect to the mean counting rates (given as numbers over each line) using a WIMP-nucleon cross section of 10$^{-8}$ pb. Calculated for $\rhodrat = 0.5$ and $M_{WIMP}$ (left to right) of 50\,\GeV, 200\,\GeV\ and 500\,\GeV. The (blue/dashed) line is the modulation signal obtained from the SHM, the (red/dot-dashed) line is the modulation signal from the dark disk and the (black/solid) line is the total modulation signal. \textit{Note the different vertical scales in each of the three plots.}}
\end{figure*}

The motion of the Earth around the Sun gives rise to an annual modulation of the event rate in a detector. The annual modulation is more pronounced for the dark disk (Fig.2), since the relative change to the mean streaming velocity owing to the Earth's motion is larger ($\sim$19\%) compared to the SHM ($\sim$6\%). The phase  (defined at maximum rate) of the dark disk and the SHM differ because the Sun's motion is slightly misaligned to the dark disk. While the phase of each component does not depend on $M_{WIMP}$, their sum does because their amplitudes depend on $M_{WIMP}$. This is a new effect introduced by the presence of the dark disk that allows $M_{WIMP}$ to be uniquely determined from the phase of the modulation signal, for given $\rhodrat$ \cite{darkdisk}.

\begin{figure*}
\begin{center}
\includegraphics[width=0.48\textwidth]{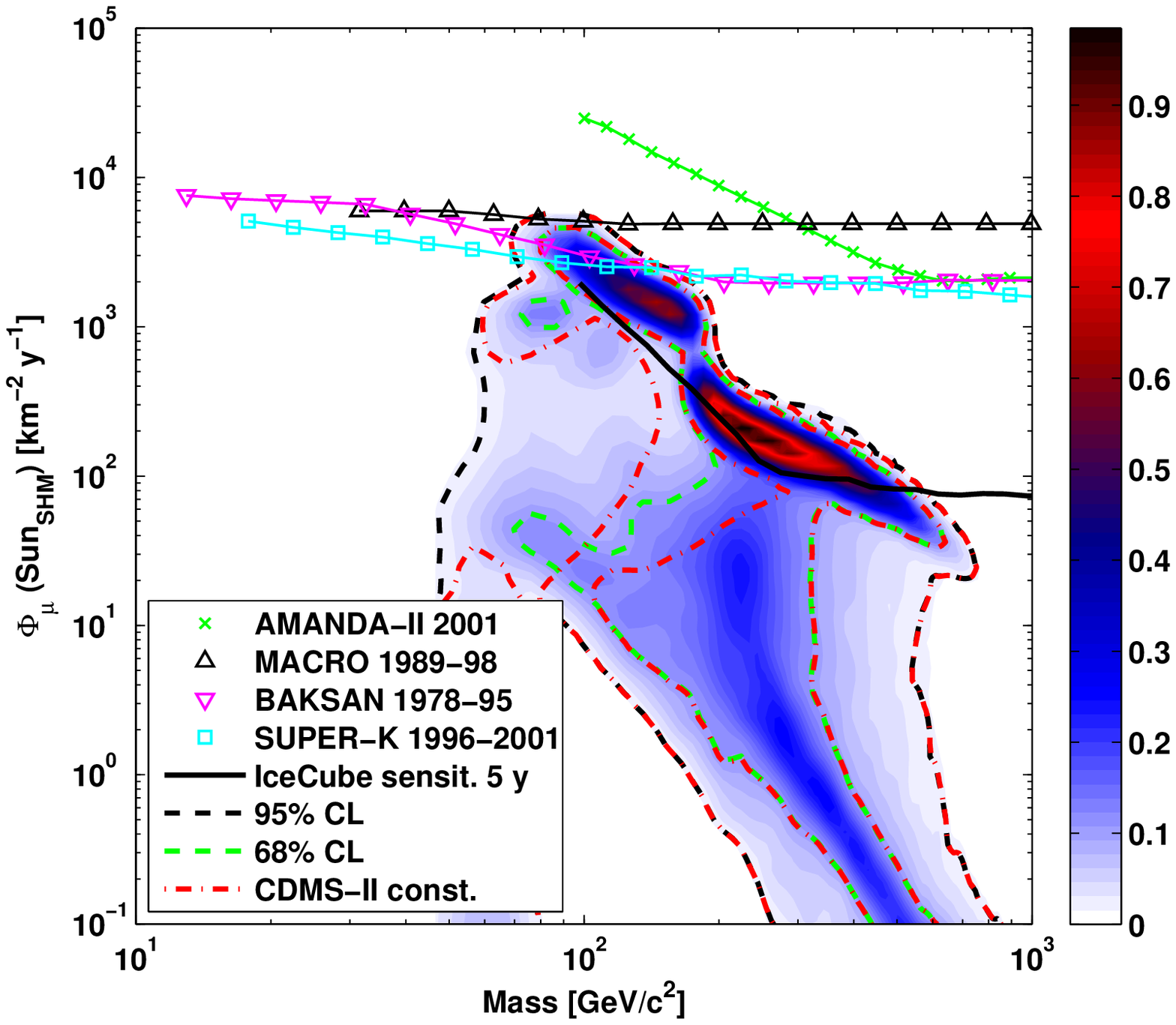}
\includegraphics[width=0.48\textwidth]{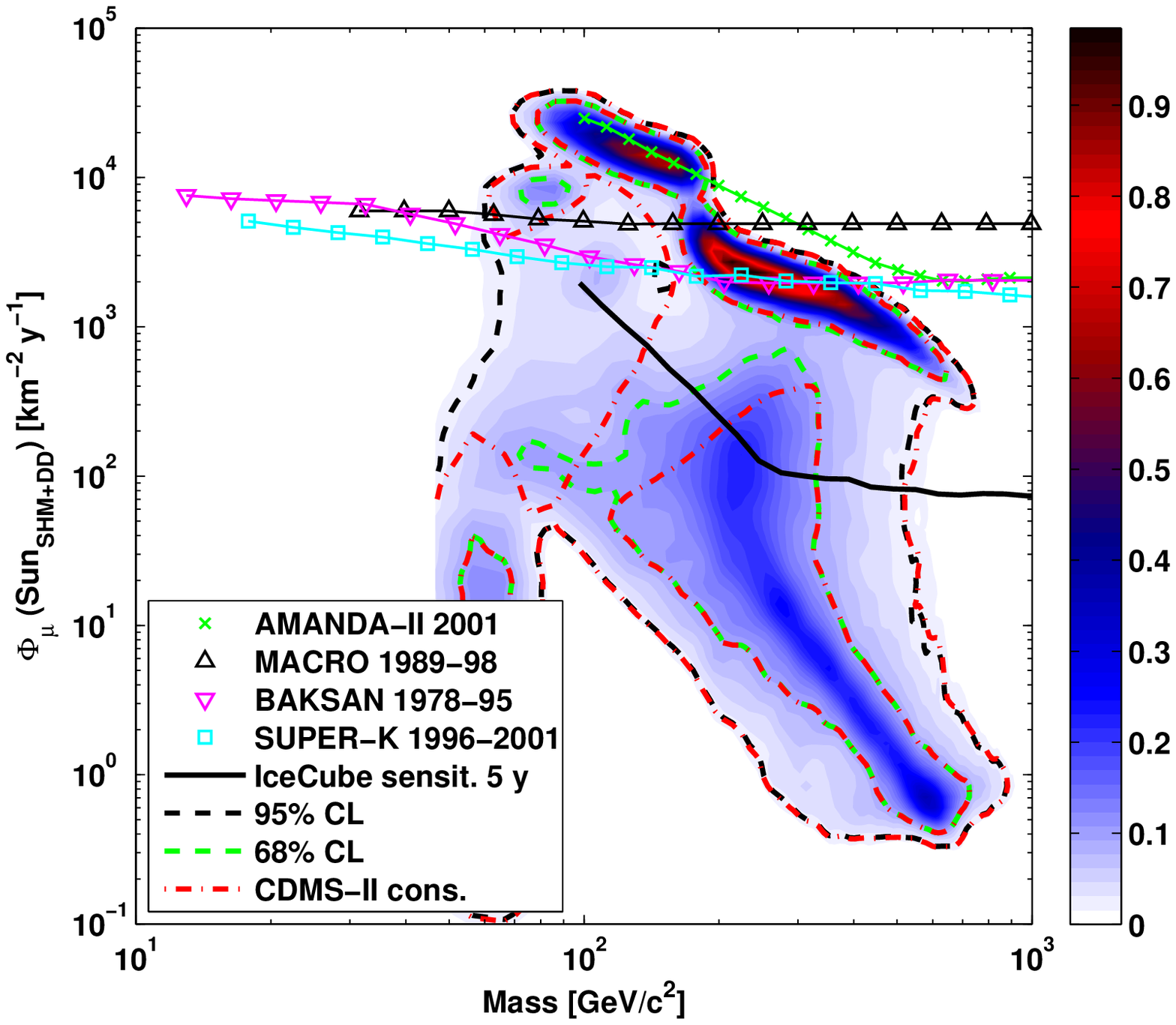}\\

\caption{Muon flux at the Earth as a function of the WIMP mass from neutrinos originated in the Sun in the SHM (left) and the summed flux of the SHM and dark disk component (right). The closed contours give the allowed regions containing 95\% (black/dashed) and 68\% (green/dashed) of the total probability. The red contours are obtained by including constraints on the spin-independent cross section from the CDMS-II experiment. The colorcode gives the relative probability density. Current experimental upper limits on the muon flux and expected sensitivities of future neutrino telescopes are shown.}
\end{center}
\end{figure*}

\section{Indirect Detection Experiments}
 Indirect detection experiments measure the decay products of WIMP annihilation. We focus here on neutrino telescopes that could measure the neutrino flux from the Sun and Earth caused by WIMP annihilation in the core regions. WIMPs streaming through the Earth and Sun lose kinetic energy by elastic scattering on nuclei, and become gravitationally captured if their velocity drops below the escape velocity. The accumulation in the core regions enhances the probability for annihilation, producing an increased flux of annihilation products, but only neutrinos will escape. Since the initial velocity of particles in the dark disk is lower compared to particles in the SHM, less scattering is needed to become gravitationally captured. The higher capture rate of particles from the dark disk directly gives rise to a higher flux of neutrinos from annihilation. These neutrinos can be detected in large area neutrino telescopes like IceCube by the back conversion of muon neutrinos to muons in the target material. 

We generated a theory sample of CMSSM models compatible with current experimental constrains using the SuperBayes code \cite{SuperBayes}. For this sample the muon flux at the Earth was calculated with the DarkSusy code \cite{DarkSusy}. The muon flux expected from the SHM is shown in Fig.3 (left panel) along with current experimental constrains and a projected sensitivity for the IceCube detector. For  $\rhodrat=0.5$ the muon flux at the Earth from neutrinos originated in the Sun is increased by a factor $\sim$5-10. The total muon flux is the sum of the SHM and the dark disk component giving the total expected muon flux shown in Fig.3 (right panel), which is dominated by the dark disk contribution. By the inclusion of the dark disk component, current experimental constraints are probing the allowed parameter space much more than in the SHM alone, and experiments like IceCube will be able to probe a large fraction of the parameter space. 

In both plots, the second (red/dot-dashed) closed contours represents the theory sample if constraints on the spin - independent cross section from the CDMS-II \cite{CDMS} experiment are imposed. This shows the high complementarity of direct and indirect detection approaches, since they probe different regions of the allowed parameter space. 

\section{Summary}
 The Milky Way's dark disk has unique features in direct detection experiments. The lower relative velocity significantly increases the differential rates at low recoil energies. For $M_{WIMP} \gtrsim 50$\, GeV, the dark disk contribution lies above current detector thresholds. The total rate in a detector is given by the sum of the two components, leading to a characteristic mass dependent shape of the differential recoil spectrum, which will improve the constraints on the particle mass upon detection. The mass dependent phase of the annual modulation signal allows a unique determination of the particle mass for known properties of the dark disk.

The lower initial velocity of particles in the dark disk gives a higher capture probability in the Sun. For  $\rhodrat=0.5$ the muon flux at the Earth from neutrinos originating in the Sun is significantly enhanced. The higher flux rises great prospects on sensitivity to allowed theoretical parameter space for current and future neutrino telescopes.

\end{document}